\PassOptionsToPackage{table}{xcolor}
\documentclass[conference]{IEEEtran}
\ifCLASSINFOpdf
\else
\fi

\usepackage[noadjust]{cite}

\usepackage{todonotes}
\usepackage{hyperref}
\usepackage{subcaption}
\usepackage{censor}
\usepackage[numbers,sort&compress]{natbib}
\usepackage{pgf-umlsd}
\usepackage{amsthm}
\theoremstyle{definition}
\newtheorem{exmp}{Example}[section]
\newcommand*\rot{\rotatebox{85}}

\usepackage{enumitem}
\usepackage{algorithm}
\usepackage{algpseudocode}
\definecolor{lightgray}{gray}{0.94}
\usepackage{cleveref}[2012/02/15]

\crefformat{footnote}{#2\footnotemark[#1]#3}

\usepackage{orcidlink}

\usepackage{listings}
\lstdefinestyle{bash}{language=bash,
    morekeywords={docker},
}
\usepackage[frozencache,cachedir=minted]{minted}
\hypersetup{
pdftitle={How Workflow Engines Should Talk to Resource Managers: A Proposal for a Common Workflow Scheduling Interface},
pdfsubject={Common Workflow Scheduler},
pdfauthor={Fabian Lehmann, Jonathan Bader, Friedrich Tschirpke, Lauritz Thamsen, and Ulf Leser},
pdfkeywords={Scientific Workflow, Scheduling, Workflow Management System, Resource Manager, Common Workflow Scheduler}
}

\newcommand\copyrighttext{%
    \footnotesize \textcopyright 2023 IEEE. Personal use of this material is permitted.
    Permission from IEEE must be obtained for all other uses, in any current or future
    media, including reprinting/republishing this material for advertising or promotional
    purposes, creating new collective works, for resale or redistribution to servers or
    lists, or reuse of any copyrighted component of this work in other works.
    DOI: \href{https://doi.org/10.1109/CCGrid57682.2023.00025}{https://doi.org/10.1109/CCGrid57682.2023.00025}}
\newcommand\copyrightnotice{%
    \begin{tikzpicture}[remember picture,overlay]
        \node[anchor=south,yshift=10pt] at (current page.south) {\fbox{\parbox{\dimexpr\textwidth-\fboxsep-\fboxrule\relax}{\copyrighttext}}};
    \end{tikzpicture}%
}

\def\BibTeX{{\rm B\kern-.05em{\sc i\kern-.025em b}\kern-.08em
    T\kern-.1667em\lower.7ex\hbox{E}\kern-.125emX}}
\begin{document}
\title{How Workflow Engines Should Talk to Resource Managers: A Proposal for a Common Workflow Scheduling Interface}
\author{

        \IEEEauthorblockN{Fabian Lehmann\orcidlink{0000-0003-0520-0792}\IEEEauthorrefmark{1}, Jonathan Bader\orcidlink{0000-0003-0391-728X}\IEEEauthorrefmark{2}, Friedrich Tschirpke\orcidlink{0000-0002-9376-2068}\IEEEauthorrefmark{1}, Lauritz Thamsen\orcidlink{0000-0003-3755-1503}\IEEEauthorrefmark{3}, and Ulf Leser\orcidlink{0000-0003-2166-9582}\IEEEauthorrefmark{1}}
        \IEEEauthorblockA{
            \IEEEauthorrefmark{1}
            \{fabian.lehmann, tschirpf, leser\}@informatik.hu-berlin.de,
            Humboldt-Universität zu Berlin, Germany
        }
        \IEEEauthorblockA{
            \IEEEauthorrefmark{2}
            jonathan.bader@tu-berlin.de,
            Technische Universität Berlin, Germany
        }
        \IEEEauthorblockA{
            \IEEEauthorrefmark{3}
            lauritz.thamsen@glasgow.ac.uk, University of Glasgow, United Kingdom
        }
  
    }
\maketitle
\copyrightnotice
\begin{abstract}
Scientific workflow management systems (SWMSs) and resource managers together ensure that tasks are scheduled on provisioned resources so that all dependencies are obeyed, and some optimization goal, such as makespan minimization, is achieved. In practice, however, there is no clear separation of scheduling responsibilities between an SWMS and a resource manager because there exists no agreed-upon separation of concerns between their different components. 
This has two consequences. 
First, the lack of a standardized API to exchange scheduling information between SWMSs and resource managers hinders portability. 
It incurs costly adaptations when a component should be replaced by a different one (e.g., an SWMS with another SWMS on the same resource manager). 
Second, due to overlapping functionalities, current installations often actually have two schedulers, both making partial scheduling decisions under incomplete information, leading to suboptimal workflow scheduling.

In this paper, we propose a simple REST interface between SWMSs and resource managers, which allows any SWMS to pass dynamic workflow information to a resource manager, enabling maximally informed scheduling decisions. 
We provide an implementation of this API as an example, using Nextflow as an SWMS and Kubernetes as a resource manager.
Our experiments with nine real-world workflows show that this strategy reduces makespan by up to 25.1\% and 10.8\% on average compared to the standard Nextflow/Kubernetes configuration.
Furthermore, a more widespread implementation of this API would enable leaner code bases, a simpler exchange of components of workflow systems, and a unified place to implement new scheduling algorithms.

\end{abstract}

\begin{IEEEkeywords}
Scientific Workflow,
Scheduling,
Workflow Management System,
Resource Manager,
Common Workflow Scheduler
\end{IEEEkeywords}

\section{Introduction}
Working with large amounts of data has become an everyday task for scientists, especially in the natural sciences.
For example, new machines allowing low-cost DNA and RNA sequencing today are available in thousands of laboratories worldwide, each able to generate between several gigabytes and multiple terabytes of data per day, building an important cornerstone for personalized medicine~\citep{yates2021reproducible, garcia2020sarek, muirRealCostSequencing2016}.
In Remote Sensing, Earth observation satellites acquire terabytes of image data daily, used for applications such as land use change detection~\citep{rs13061125, Sudmanns_Tiede_Augustin_Lang_2019, lehmannFORCENextflowScalable2021}.
For instance, in 2021, the European Space Agency's Sentinel mission of the EU's Copernicus program generated 7.34PiB of new data~\citep{Castriotta_2022}.

\begin{figure}
    \centering
    \begin{subfigure}{.14\columnwidth}
      \centering
      \includegraphics[width=.7\columnwidth,trim={23mm 120mm 285mm 5mm},clip]{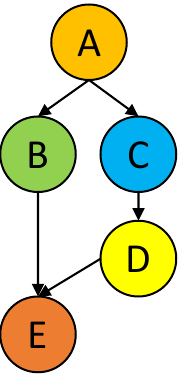}
      \caption{Abstract DAG}
	  \label{fig:exampleAbstractDAG}
    \end{subfigure}%
    \begin{subfigure}{.30\columnwidth}
      \centering
      \includegraphics[width=.7\columnwidth,trim={9mm 104mm 270mm 5mm},clip]{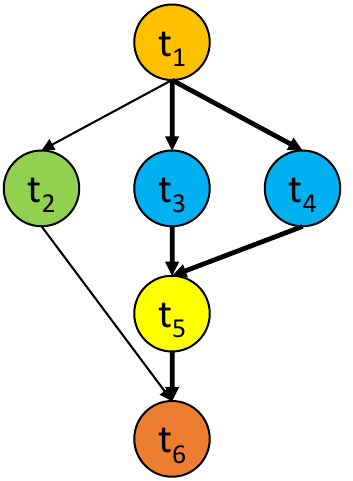}
      \caption{Physical DAG}
	  \label{fig:exampleDAG}
    \end{subfigure}%
    \begin{subfigure}{.22\columnwidth}
      \centering
      \includegraphics[width=.7\columnwidth,trim={4mm 119mm 304mm 5mm},clip]{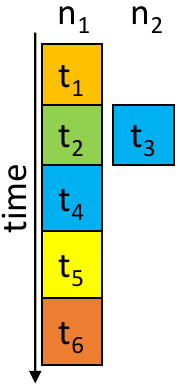}
      \caption{Schedule 1}
	  \label{fig:schedule1}
    \end{subfigure}
    \begin{subfigure}{.22\columnwidth}
      \centering
      \includegraphics[width=.7\columnwidth,trim={4mm 119mm 304mm 5mm},clip]{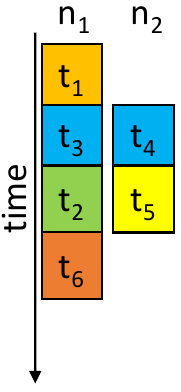}
      \caption{Schedule 2}
	  \label{fig:schedule2}
    \end{subfigure}
    \caption{An example of an abstract DAG and a belonging physical DAG instance with six tasks and seven edges; the critical path is bold; two possible schedule plans to execute the DAG on two nodes.}
    \label{fig::dagAndSchedule}
\end{figure}

Natural scientists process such data using pipelines of independently developed tools.
Scientific workflow management systems (SWMSs), such as Nextflow~\citep{ditommasoNextflowEnablesReproducible2017}, Snakemake~\citep{koesterSnakemake2012}, or Pegasus~\citep{deelman2019evolution}, are software infrastructures that help to organize the interplay and distributed execution of these individual analysis steps (also called tasks)~\citep{liuSurveyDataIntensiveScientific2015, wcs2021community}.
They typically require the analysis to be described in the form of workflows, where the dependencies between the individual tasks are defined as abstract directed acyclic graphs (DAGs).
Figure~\ref{fig:exampleAbstractDAG} shows an example of such an abstract DAG with five abstract tasks represented as nodes and five dependencies represented as edges.
To execute a workflow, an SWMS compiles the abstract DAG into a physical DAG, controls all data dependencies, and generates prioritized lists of ready-to-run tasks.
Figure~\ref{fig:exampleDAG} shows the final physical DAG with six concrete tasks and seven edges.

The lists of tasks are typically communicated to a resource manager, such as Kubernetes~\citep{kubernetes}, HTCondor~\citep{condor}, or Slurm~\citep{slurm}, which schedules them across the set of available nodes, taking into account specific task requirements and the current node load~\citep{topcuogluPerformanceeffectiveLowcomplexityTask2002}.

However, this setup has considerable drawbacks.
First, the SWMS lacks the resource manager's knowledge of available resources and, thus, cannot prioritize tasks optimally.
On the other hand, the resource manager lacks the SWMS's knowledge of future tasks and, thus, cannot prioritize tasks optimally according to their dependent future tasks.
This leads to suboptimal scheduling decisions, such as placing a task on the critical path on a slow machine simply because the critical path is unknown to the resource manager.

\begin{exmp}
Assume the physical DAG from Figure~\ref{fig:exampleDAG} is scheduled on a two-node cluster.
The SWMS does not know about a task until all predecessor tasks have been finished.
For simplicity, assume that all tasks take one time unit on both Node $n_1$ and Node $n_2$.
The critical path of the workflow is shown in bold.
When starting the workflow, the SWMS first sends $t_1$ to the resource manager, which places the task on either $n_1$ or $n_2$ at the first time unit.
After $t_1$ is completed, the SWMS determines that $t_2$, $t_3$, and $t_4$ are ready for execution and sends them to the resource manager.
Since none of the three tasks has been finished yet, the SWMS does not know that $t_5$ will exist.
Without knowledge of future tasks, the SWMS and the resource manager must treat all three tasks equally.
For example, it may schedule $t_2$ and $t_3$ first based on a FIFO policy.
At the third time unit, the resource manager would schedule $t_4$; when $t_4$ has finished, $t_5$ is submitted and scheduled.
Finally, $t_6$ is submitted and runs on either node $n_1$ or node $n_2$.
Figure~\ref{fig:schedule1} shows the tasks running in parallel.
The execution takes five time units.

In contrast, the resource manager could make more informed decisions knowing future tasks and their dependencies.
After $t_1$ has finished, it could then prioritize and schedule $t_3$ and $t_4$ in the second step since they are on the critical path and the scheduler is aware of upcoming tasks from the abstract task D.
Figure~\ref{fig:schedule2} depicts this run, which takes only four units of time.
\end{exmp}

Today, knowledge exchange between an SWMS and a resource manager is difficult to achieve because there is no standardized API between both components.
As a result, many resource managers do not implement schedulers that are able to consider future and dependent tasks~\citep{rodrigoEnablingWorkflowAwareScheduling2017}.

Further, even resource managers that can take dependency structures into account, such as Slurm~\citep{slurm} or HTCondor/DAGMan~\citep{couvaresWorkflowManagementCondor2007}, can only do so if they know the entire DAG at the workflow's start time.
This, however, is not the case for dynamic workflows as supported by modern SWMSs such as Nextflow or Parsl~\citep{babujiParslScalableParallel2019}.
In dynamic workflows, the structure of the DAG is data dependent; for example, the choice between two different branches within the DAG may depend on a property of the input data (conditional), or the breadth of scatter operations may depend on the size of some intermediate data file.
Besides producing suboptimal schedules, the lack of a common API also implies that developers of scheduling algorithms need to implement their ideas anew for every combination of SWMS and resource manager, which might be one of the reasons for the rather slow uptake of new scheduling ideas in practice.

In this paper, we present a proposal for a REST-API between an SWMS and a resource manager, allowing the resource manager to become the only place for taking optimally informed scheduling decisions while freeing the SWMS from any such considerations.
We designed the API for SWMSs that create DAGs dynamically by always conveying all sure future DAG parts to the resource manager for consideration, but it also works with static DAGs.
We implemented a prototype for the combination of Nextflow and Kubernetes and showed that it consistently produces faster schedules than the current standard configuration across a number of different real-world workflows.
Specifically, our paper makes the following contributions:
\begin{itemize}
    \item We propose a simple REST API for the interface between an SWMS and a resource manager to dynamically inform the resource manager about future tasks and their relationship to the currently ready-to-run tasks. 
    \item We provide an open-source prototype implementation of this API for Nextflow\footnote{\label{git::prototypeNxf}\url{https://github.com/CommonWorkflowScheduler/Nextflow}} and Kubernetes\footnote{\label{git::prototypeK8s}\url{https://github.com/CommonWorkflowScheduler/KubernetesScheduler}}.
    Note that replacing, for example, Nextflow with Airflow in this configuration would only require adapting Airflow to this API, while the Kubernetes site could remain unchanged.
    \item We implemented a total of 21 scheduling algorithms as an extension to Kubernetes that can take advantage of the information available through the API.
    \item We evaluated our prototype and all scheduling algorithms using nine nf-core~\citep{ewelsNfcoreFrameworkCommunitycurated2020} workflows, with a total runtime of approximately eleven days.
    All nf-core workflows are executed dynamically.
    Our results show that the informed scheduling strategies yield more efficient schedules compared to the standard configuration, with Rank (Min) Round-robin performing best on average.
    \item We make all the code, the experimental setup, and results freely available via GitHub\footnote{\label{git::experiments}\url{https://github.com/CommonWorkflowScheduler/ExperimentsAndResults}}.
    \item We developed a plugin\footnote{\url{https://github.com/CommonWorkflowScheduler/nf-cws}} for Nextflow based on our prototype that enables the use of the Common Workflow Scheduler with the official Nextflow version.
\end{itemize}

\section{Background}
\begin{figure}[b!]
  \centering
  \includegraphics[width=\columnwidth,trim={2.8mm 139.0mm 228.7mm 7.2mm},clip]{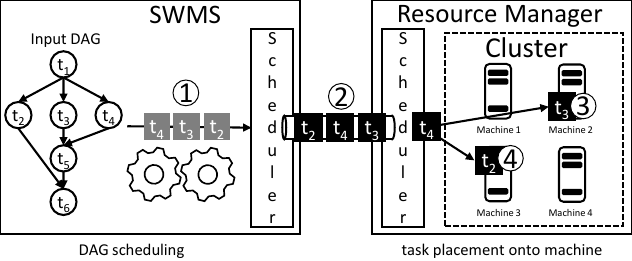}
  \caption{Architecture overview: two schedulers for the state-of-the-art interaction between SWMSs and resource managers}
  \label{fig:overview}
\end{figure}%
In this section, we illustrate the scheduling model of scientific workflows.
We also explain the scheduling component of the SWMS and the resource manager.

\subsection{Interaction between SWMS and Resource Manager}
This section describes the interaction between an SWMS and a resource manager, as performed by state-of-the-art systems such as Nextflow, Snakemake, Airflow, and Argo, which submit tasks to resource managers, e.g., Slurm, Kubernetes, or HTCondor.
Figure~\ref{fig:overview} provides an overview of the execution environment.
The figure's left shows the SWMS with a workflow represented as a DAG.
On the right side, we see the resource manager abstracting the cluster and its nodes.
The figure shows the queue between the SWMS and the resource manager that the resource manager exposes to receive tasks.

Nodes in a cluster are often heterogeneous~\citep{hutson2019managing,cpuperformance}.
Heterogeneity can mean the same CPU architecture with a different clock speed, a CPU from a different generation or manufacturer with the same number of cores, or a different generation of memory.
These characteristics must be taken into account during scheduling because even nodes with the same number of CPU cores or the same amount of memory can produce completely different task runtimes; however, the resource manager abstracts detailed hardware information.

\paragraph{SWMS's scheduling}
The SWMS is aware of the abstract DAG's structure at any time, i.e., it knows the task dependencies and the overall workflow structure.
However, not all systems initially know the entire physical DAG.
For example, SWMSs such as Nextflow and Parsl have dynamic DAGs, and physical instances are created at runtime depending on the previous results.
Accordingly, it is impossible to do static scheduling at the workflow's start.
Most SWMSs come with their scheduler to prioritize tasks based on the DAG, as shown in Figure~\ref{fig:overview}.
Other SWMSs, such as Nextflow, submit a task as soon as it becomes known without prioritizing it.

\begin{exmp}
We use the example from Figure~\ref{fig::dagAndSchedule} to follow the data flow in Figure~\ref{fig:overview}.
Internally, at the second time unit, $t_2$, $t_3$, and $t_4$ are ready-to-run and are passed into the SWMS's scheduler~\textcircled{\raisebox{-0.9pt}{1}}.
The SWMS's scheduler then considers the tasks' dependencies and requirements to prioritize the tasks and submit them to the resource manager's queue~\textcircled{\raisebox{-0.9pt}{2}}.
The SWMS does not know the number of nodes in the cluster and the current resources available on each node.
Accordingly, if we assume that they can all start in parallel, all three tasks $t_2$, $t_3$, and $t_4$ can be considered equal.
With more than two nodes, the tasks in the DAG in Figure~\ref{fig:exampleDAG} can always finish within four time units independent of the task order.
Since the SWMS does not know the number of nodes, their attributes, and their occupation, the SWMS only decides when to submit a task but not where to run it.
\end{exmp}

\paragraph{Resource manager's scheduling}
The resource manager's scheduler then handles the queued tasks independently. 
However, some resource managers support dependencies so that a task will not start until all its predecessors have finished~\citep{rodrigoEnablingWorkflowAwareScheduling2017}.
While the SWMS has submitted the tasks in the order $t_3$, $t_4$, and $t_2$~\textcircled{\raisebox{-0.9pt}{2}}, the resource manager cannot guarantee the schedule plan, as shown in Figure~\ref{fig:schedule2}.
For example, the resource manager could reorder the tasks.
In~\textcircled{\raisebox{-0.9pt}{3}}, the resource manager starts $t_3$ first, and in~\textcircled{\raisebox{-0.9pt}{4}}, it starts $t_2$; meanwhile, $t_4$ is still waiting to be executed.
This can happen if tasks have different requirements and there are not enough unallocated resources on a node for $t_4$ but $t_2$ at the time of scheduling.

\subsection{Kubernetes}
Here, we explain the concepts of Kubernetes as it is necessary to follow our prototype implementation presented in Section~\ref{chap::impl}.
Kubernetes\footnote{\url{https://kubernetes.io/}} is an open-source resource manager for orchestrating large clusters of nodes~\citep{burnsBorgOmegaKubernetes2016} and is used by many of today's SWMSs such as Nextflow, Argo, and Snakemake.
Kubernetes provides a unified view over all managed nodes, offering an interface to submit, for example, single tasks or deployments with automatic scaling.

Kubernetes is built around the concept of pods.
A pod wraps one or more containers that Kubernetes executes together.
In the workflow domain, tasks are expressed as pods, mostly using a single container.

The Kubernetes resource manager follows a declarative paradigm. 
Accordingly, all its components write the desired state into a database, and Kubernetes tries to achieve it.
The use of the declarative paradigm makes extensions and replacements of Kubernetes components relatively easy, as only the desired state needs to be defined.

\subsection{Nextflow}

Nextflow~\citep{ditommasoNextflowEnablesReproducible2017} is an open-source SWMS with a Groovy-like DSL.
We use Nextflow as SWMS in our prototype in Section~\ref{chap::impl}.
Nextflow aims to enable reproducible and scalable workflows in bioinformatics by using lightweight virtualization platforms such as Docker.
However, Nextflow also gains increasing interest from other domains, such as remote sensing~\citep{lehmannFORCENextflowScalable2021}.
Nextflow allows a workflow to be defined independently of the execution environment and enables the workflows to run on many resource managers, such as Kubernetes, Slurm, HTCondor, and cloud providers, such as Google and AWS.

The workflow execution in Nextflow is data-flow-driven~\citep{tommasoQuickOverviewNextflow2022}.
Accordingly, all data exchanged between tasks is put into channels.
Channels allow the automatic determination of parallelizable tasks.
In addition, Nextflow determines tasks at runtime whenever data is in a channel.

To work with all the different resource managers, Nextflow abstracts them from the user.
Nextflow submits every ready-to-run task individually for execution.
With Kubernetes, each task is executed as a pod.
However, Nextflow neither applies task ordering nor task placement.
Accordingly, Nextflow could benefit from workflow-aware scheduling.

\section{Related Work}
In this section, we will provide an overview of the related work.
First, scheduling algorithms and, second, SWMS scheduling to resource managers.

\subsection{Workflow Scheduling Algorithms}
Many scheduling algorithms exist for scientific workflows.
A well-known representative of workflow scheduling is the Heterogeneous Earliest-Finish-Time (HEFT) algorithm~\citep{topcuogluPerformanceeffectiveLowcomplexityTask2002}.
HEFT consists of two stages.
In the first stage, HEFT prioritizes tasks by calculating the average remaining computation and communication time between a task and the workflow's end.
In the second stage, all tasks are assigned to nodes, starting with the highest-ranked task.
Many HEFT adaptations exist, such as BDHEFT~\cite{Verma_Kaushal_2015} and DQ-HEFT~\citep{kaurDeepQLearningbasedHeterogeneous2022}.

Other scheduling approaches deal with wrong decisions and readjust the scheduling plan.
For example, AHEFT~\citep{yuAdaptiveReschedulingStrategy2007} recalculates the scheduling plan using HEFT once a task fails, new resources become available, or the performance varies. 
However, it still requires the full physical DAG and accurate runtime predictions.

P-HEFT~\citep{barbosa2011dynamic} takes into account the dynamic behavior of workflows and deals with unknown task arrival times.
Therefore, only the abstract DAG needs to be known a priori to compute the longest path between a task and an exit task and to prioritize the tasks accordingly.

Min-Min and Max-Min only consider a batch of ready-to-run tasks~\citep{maheswaranDynamicMappingClass1999}.
Min-Min schedules the task with the earliest expected finish time to the node where it is expected to finish the earliest.
In contrast, the Max-Min strategy schedules the task with the latest expected completion time to the node where it will complete the fastest.

Crucially, these theoretical scheduling methods assume global knowledge, which makes them challenging to implement in SWMSs that rely on resource managers~\citep{schwiegelshohnHowDesignJob2015}.
As a result, these methods have been presented mainly using simulations.
They require the knowledge that is currently distributed between two schedulers.
In our approach, we propose to use a single scheduler that allows knowledge transfer for a more comprehensive view.
Our scheduler API is designed to support both static and dynamic DAGs.
We enable the SWMS to submit the entire physical DAG at the beginning or to generate it dynamically.

\subsection{Scheduling of SWMS Tasks on Resource Managers}

\citeauthor{shanContainerizedWorkflowBuilder2021} develop a strategy to use Kubernetes native functions to schedule workflow tasks~\citep{shanContainerizedWorkflowBuilder2021}.
The authors claim a faster startup time because no SWMS is involved.
They wrap a workflow DAG into tasks and their dependencies, and Kubernetes starts the successor tasks when all predecessor tasks have succeeded.
However, compared to our approach, they do not address task placement or ordering and require the complete DAG in advance.
Also, our API-based approach works for other SWMSs and is not limited to Kubernetes.

Slurm~\cite{slurm} supports task dependencies out of the box.
Again, this only works for static workflows where all tasks are initially submitted.
\citeauthor{rodrigoEnablingWorkflowAwareScheduling2017} address the problem that Slurm does not start prioritizing tasks until all dependencies are satisfied by reordering the submission queue~\citep{rodrigoEnablingWorkflowAwareScheduling2017}.
Their approach only addresses task prioritization but not placement.

Flux~\cite{ahnFluxOvercomingScheduling2020} is a resource manager primarily designed for exascale workflows.
Therefore, it uses hierarchical scheduling, which supports plugins for specific scheduling strategies at different hierarchy levels.
As with all resource managers, an SWMS needs to be specifically adapted for Flux.

While some particular resource managers deal with dependencies, other resource managers, such as Kubernetes, assume that each task is independent of other tasks~\citep{rodrigoEnablingWorkflowAwareScheduling2017}.

Tarema~\citep{baderTarema2021} profiles the target infrastructure and clusters nodes into groups according to their hardware performance. 
The SWMS then labels tasks according to their resource utilization characteristics.
Based on these labels, Tarema evenly maps tasks to node groups.
Tarema has been implemented and tested using Nextflow and a custom Kubernetes scheduler.
Unlike our approach, Tarema does not consider the workflow DAG and treats tasks independently. 
Also, while Tarema is a prototype, our API-based approach is more general.

\citeauthor{buxHiWAYExecutionScientific2017} extend Hadoop YARN with their workflow scheduler HI-WAY~\citep{buxHiWAYExecutionScientific2017}.
HI-WAY handles DAGMan, Galaxy, and Cuneiform workflows and provides various scheduling algorithms such as HEFT.
To perform HEFT scheduling, HI-WAY uses historical provenance data stored in its own provenance manager.
The scheduling plan is adjusted as soon as new measurements become available.
Again, this approach requires static DAGs.

DAGMan~\citep{couvaresWorkflowManagementCondor2007} receives a workflow definition that can be generated by different SWMSs and submits the tasks to HTCondor.
DAGMan ensures that the workflow's dependencies are kept by observing the tasks' state.
Accordingly, DAGMan takes over the entire workflow execution, excluding SWMSs from workflow execution.
Therefore, DAGMan needs static DAGs and submits each task to HTCondor as soon as all dependencies are fulfilled.
For example, Pegasus~\citep{deelmanPegasusWorkflowManagement2015} splits hierarchical workflows into smaller sub-workflows and submits them to DAGMan for execution~\citep{deelmanPegasusDAGManConcept}.
Pegasus adds priorities to the tasks to influence the order of execution.

DAGwoman~\citep{tschagerDAGwomanEnablingDAGManlike2012} is a standalone tool in the userspace that handles DAGMan workflow definitions but executes them on resource managers other than HTCondor.

State-of-the-art SWMSs such as Nextflow~\citep{ditommasoNextflowEnablesReproducible2017}, Snakemake~\citep{koesterSnakemake2012}, Argo, and Airflow use the scheduler of a resource manager~\citep{shanContainerizedWorkflowBuilder2021} and at most do some task prioritization.

Unlike our method, the existing schedulers are tightly coupled between a specific resource manager and a specific SWMS.
In addition, most approaches only work for static DAGs.

\section{Scheduler API}
In this section, we will explain what information a resource manager's scheduler needs to make good scheduling decisions.
Afterward, we will design a REST-API to exchange this information.

\subsection{API Requirements}
\begin{figure}
  \centering
  \includegraphics[width=\columnwidth,trim={2.8mm 139.0mm 228.7mm 7.2mm},clip]{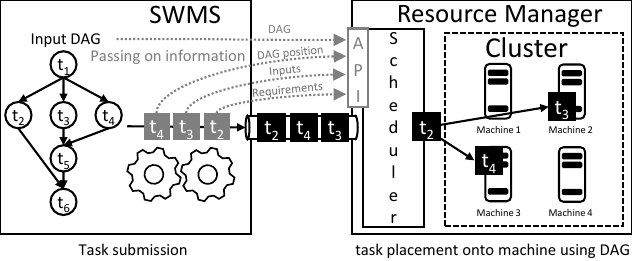}
  \caption{One workflow-aware scheduler with comprehensive knowledge; information transfer using our API}
  \label{fig:overviewMyApproach}
\end{figure}%
Figure~\ref{fig:overviewMyApproach} shows an architectural overview of our proposed system.
We have a single scheduler running on the resource manager's side that receives all missing information from the SWMS.
With this information, the scheduler can make more advanced scheduling decisions than before.

First, we transfer the DAG in the available form.
The DAG may change during execution due to conditioning.
For example, new tasks may be added to or removed from the DAG.

Next, the scheduler needs to know the physical tasks - concrete tasks that the resource manager needs to execute and that the resource manager's scheduler needs to assign to a node.
While some SWMSs know all the concrete tasks at the start, other SWMSs only know the tasks that are ready to run.
In the first case, the scheduler must know the dependencies between the concrete tasks.
In the second case, a task can start as soon as it is submitted.
In addition, the SWMS can withdraw a physical task that was not ready to run due to a condition evaluated later.

The scheduler considers all tasks that have already been submitted and, thus, might already assign a task to a node right before a more suitable task arrives.
Therefore, we need a batching mechanism to submit multiple ready-to-run tasks simultaneously. 

In addition, the scheduler needs a link between physical and abstract tasks to detect similar tasks and apply knowledge of previous task executions to them.
The scheduler can also leverage a connection between resubmitted physical tasks or task instances based on the same abstract task for advanced scheduling.
Further, the scheduler must receive the user's annotation for CPU, memory, or runtime requirements.
In addition, a task's input and expected output files will help the workflow scheduler achieve better task placement~\citep{brykStorageawareAlgorithmsScheduling2016, donnellyConfugaScalableData2015}.

Not only does the resource manager's scheduler need to receive information, but the SWMS might request statistics and progress from the scheduler.
Moreover, we can imagine that the resource manager's scheduler learns task characteristics and provides them to the SWMS, overriding imprecise user annotations.
For example, we can learn a task's memory requirements~\cite{witt2019feedback,tovarJobSizingStrategy2018} or the resource manager's scheduler might limit the number of CPUs assigned to a task due to capacity constraints.

\subsection{API Design}%

\begingroup 
\setlength{\tabcolsep}{5.4mm}
\begin{table}[t]
    \centering
    \rowcolors{2}{white}{lightgray}
    \caption{Scheduler's REST-API design}
    \begin{tabular}{r|l|c}
      \# & Resource & Method \\
       \hline
      1 & /\{version\}/\{execution\} & POST \\
      2 & /\{version\}/\{execution\} & DELETE \\
      3 & /\{version\}/\{execution\}/DAG/vertices & POST \\
      4 & /\{version\}/\{execution\}/DAG/vertices & DELETE \\
      5 & /\{version\}/\{execution\}/DAG/edges & POST \\
      6 & /\{version\}/\{execution\}/DAG/edges & DELETE \\
      7 & /\{version\}/\{execution\}/startBatch & PUT \\
      8 & /\{version\}/\{execution\}/endBatch & PUT \\
      9 & /\{version\}/\{execution\}/task/\{id\} & POST \\
     10 & /\{version\}/\{execution\}/task/\{id\} & GET \\
     11 & /\{version\}/\{execution\}/task/\{id\} & DELETE \\
       \hline
    \end{tabular}
    \label{tab:rest_api}
\end{table}%
\endgroup%
\begin{algorithm}
	\caption{API Interaction from SWMS Perspective}
        \label{alg:apiInteraction}
        \small
	\begin{algorithmic}[1]
	    \State Register execution at scheduler (1)
            \State Send request: submit all vertices (3)
            \State Send request: submit all edges (5)
     
	    \While{The workflow has not yet finished}
                \If{Create abstract task}
		        \State Send request: add vertices (3)
		    \EndIf
                \If{New connection between two abstract tasks}
		        \State Send request: add edges (5)
		    \EndIf
    		\If{Withdrawn abstract task}
		        \State Send request: remove vertices (4)
		    \EndIf
                \If{Withdrawn connection between two abstract tasks}
		        \State Send request: remove edges (6)
		    \EndIf
                \If{Withdrawn physical task instance}
		        \State Send request: remove task (11)
		    \EndIf
                \If{Physical tasks are ready for execution}
		        \State Send request: start batch (7)
                    \For {all ready-to-run-tasks}
    		          \State Send request: submit task (9)
    		      \EndFor
		        \State Send request: end batch (8)
		    \EndIf
                \For {All submitted tasks}
        		  \State Send request: request state (10)
    		\EndFor
		\EndWhile
	    \State Delete execution at scheduler (2)
	\end{algorithmic} 
\end{algorithm}%
We propose an API between the resource manager and the SWMS in the form of a REST-API. 
We chose REST because any programming language used for SWMSs should support REST.
Also, using REST also makes the implementation on the resource manager's side language independent.
Table~\ref{tab:rest_api} shows the REST-API's resources, access methods, and reference number.
Algorithm~\ref{alg:apiInteraction} illustrates a typical interaction with our API during workflow execution.
Each call is referenced by its reference number.
The REST-API provides twelve resources and supports versioning.
The user or SWMS defines the \emph{execution} as a unique identifier of the current workflow execution.

To start a workflow execution, the SWMS calls the REST-API, registers a new workflow execution, and defines the scheduling strategy~(1).
Any additional information is wrapped in the request body.
After the workflow has finished successfully, because of a failure, or because of a user's interruption, the SWMS deletes the execution~(2).

After the workflow is registered, the SWMS submits the abstract DAG, including a list of abstract processes~(3), the vertices in a DAG, and their dependencies - the edges~(5).
Due to conditioned executions, the SWMS can delete abstract processes~(4) and edges~(6) during execution.

To submit tasks, the SWMS can optionally start a batch~(7). 
Then, the resource manager will not start tasks from this batch until the SWMS closes the batch~(8).
If the SWMS has not opened a batch, the batch size is one.

An SWMS can submit physical tasks whenever a batch is open or whenever no batching is used~(9).
The SWMS defines an \emph{id} for each task.
In addition, the request includes input and output files if known.
The SWMS also sends the user-defined CPU, memory, and runtime requirements.
The REST-API returns the CPU, memory, and runtime it will use for the task.
The SWMS can query the state of a task by its \emph{id}~(10).
Finally, when a task is withdrawn, the SWMS can remove it~(11).

\section{Exemplary implementation}\label{chap::impl}
In this section, we will provide an overview of our reference implementation.
We implemented our scheduler API for Kubernetes as a state-of-the-art resource manager.
For the SWMS, we chose Nextflow as a well-known SWMS.
We refer the reader to the Appendix, where we briefly introduce how to build and use the two tools.
\subsection{Kubernetes Extension}

Initially, the SWMS announces a task using the REST-API and submits the concrete task to Kubernetes as a pod.
The task is added to an internal queue of unscheduled tasks.

Our Kubernetes scheduler runs in a standard pod next to the original Kubernetes scheduler and only acts for pods that explicitly request it.
We implemented the scheduler in Java using Fabric8\footnote{\url{https://github.com/fabric8io/kubernetes-client}} to access the Kubernetes API, SpringBoot\footnote{\url{https://spring.io/projects/spring-boot}} to expose the REST-API, and OpenAPI\footnote{\url{https://www.swagger.io/}} to document the REST-API and guide SWMS developers on how to adapt the REST-API.

Using Fabric8 components, we subscribe to the Kubernetes API, watch for changes, and define the target state.
Tracking node and pod changes gives us a global view of the cluster.
Thus, the scheduler is aware of available and occupied resources.

The scheduler provides an interface for developers to implement different scheduling strategies.
We provide 21 out-of-the-box scheduling strategies in our scheduler, that are easily extendable. 
The scheduler aligns the tasks with the abstract process in the DAG to get the rank of a task.
For extensibility, the Scheduler class keeps track of the resource usage in the cluster and provides methods that can be overwritten to handle events in the cluster, such as unreachable nodes and new pods.
The Scheduler class has an abstract method to create a task-node alignment. 
Our scheduler prototype can be found on GitHub\cref{git::prototypeK8s}.

\subsection{Nextflow Extension}
Nextflow supports Kubernetes out of the box.
Accordingly, we only add the communication to our scheduling API but no new task submission logic.
We adapted and changed the following aspects.

First, we submit the abstract DAG before any task is submitted.
We then capture any changes to the DAG and immediately pass them to our scheduler.

Second, Nextflow permanently loops over the list of ready-to-run but unsubmitted tasks.
At the beginning of this loop, we start a new batch which we close at the end of the loop.
The batch size is configurable using the Nextflow configuration.
Batching prevents the scheduler from starting a ready-to-run task on a node while a more suitable task arrives at the next moment.

Third, we submit our task to the scheduler.
We include the path to the input files, the user-defined CPU and memory requests, and the pod name.
The scheduler uses the pod name to match the submitted task with a submitted pod.

Finally, we wrap the task into a pod.
In addition, we define our new scheduler as the responsible scheduler for the pod.

Our adjusted Nextflow version can be found on GitHub\cref{git::prototypeNxf}.

\section{Evaluation}
\begingroup 
\setlength{\tabcolsep}{1.7mm}
\begin{table*} \centering
    \rowcolors{2}{white}{lightgray}
    \caption{Key workflow characteristics and experiment overview on the nine evaluation workflows}
    \begin{tabular}{c|r|r|c|r|r|r|r|r|r}
\multicolumn{1}{p{1.2cm}|}{\parbox[c][4em]{\hsize}{\centering Workflow}} & \multicolumn{1}{p{1.2cm}|}{\parbox[c][4em]{\hsize}{\centering \# tasks instances}} & \multicolumn{1}{p{1.5cm}|}{\parbox[c][4em]{\hsize}{\centering generated data}} & \multicolumn{1}{p{2cm}|}{\parbox[c][4em]{\hsize}{\centering Strategy with the best median run}} & \multicolumn{1}{p{1.8cm}|}{\parbox[c][4em]{\hsize}{\centering Original median runtime}} & \multicolumn{1}{p{1.2cm}|}{\parbox[c][4em]{\hsize}{\centering Avg. task runtime}} & \multicolumn{1}{p{1.2cm}|}{\parbox[c][4em]{\hsize}{\centering Median task runtime}}  & \multicolumn{1}{p{1.2cm}|}{\parbox[c][4em]{\hsize}{\centering Standard dev. task runtime}} & \multicolumn{1}{p{0.9cm}|}{\parbox[c][4em]{\hsize}{\centering Best median runtime}} & \multicolumn{1}{p{1.6cm}}{\parbox[c][4em]{\hsize}{\centering Improvement}} \\
\hline
RNA-Seq & 415 & 495.6 MB & Rank (Min)-Fair & 839.5s & 3.2s & 1.0s & 10.2s & 629.1s & 25.1\% \\
Sarek & 110 & 536.1 MB & Size Asc-Fair & 2,121.9s & 17.8s & 1.0s & 158.6s & 2,027.5s & 4.4\% \\
ChiP-Seq & 587 & 2,636.4 MB & Rank (FIFO)-Ran & 821.4s & 3.1s & 1.0s & 6.5s & 725.2s & 11.7\% \\
ATAC-seq & 481 & 5,790.2 MB & Rank (Min)-Ran & 867.0s & 5.5s & 2.8s & 8.9s & 748.8s & 13.6\% \\
MAG & 1,115 & 18,557.5 MB & Rank (Min)-RR & 1,254.4s & 5.7s & 2.0s & 13.1s & 1,091.6s & 13.0\% \\
AmpliSeq & 139 & 267.5 MB & Ran-Ran & 1,042.7s & 6.6s & 4.6s & 8.2s & 848.1s & 18.7\% \\
NanoSeq & 600 & 14,613.8 MB & FIFO-Ran & 809.9s & 2.7s & 0.0s & 5.5s & 747.8s & 7.7\% \\
Viralrecon & 681 & 894.1 MB & Rank (Min)-Fair & 916.4s & 2.7s & 0.1s & 8.3s & 783.5s & 14.5\% \\
Eager & 646 & 2,383.8 MB & Rank (Min)-Ran & 708.9s & 3.3s & 3.2s & 2.8s & 684.5s & 3.5\% \\
\hline
\end{tabular}
    \label{tab::strategy_runtimes}
\end{table*}
\endgroup
\begin{figure*}[t]
  \centering
  \includegraphics[width=\textwidth,trim={0.0mm 0mm 0mm 0mm},clip]{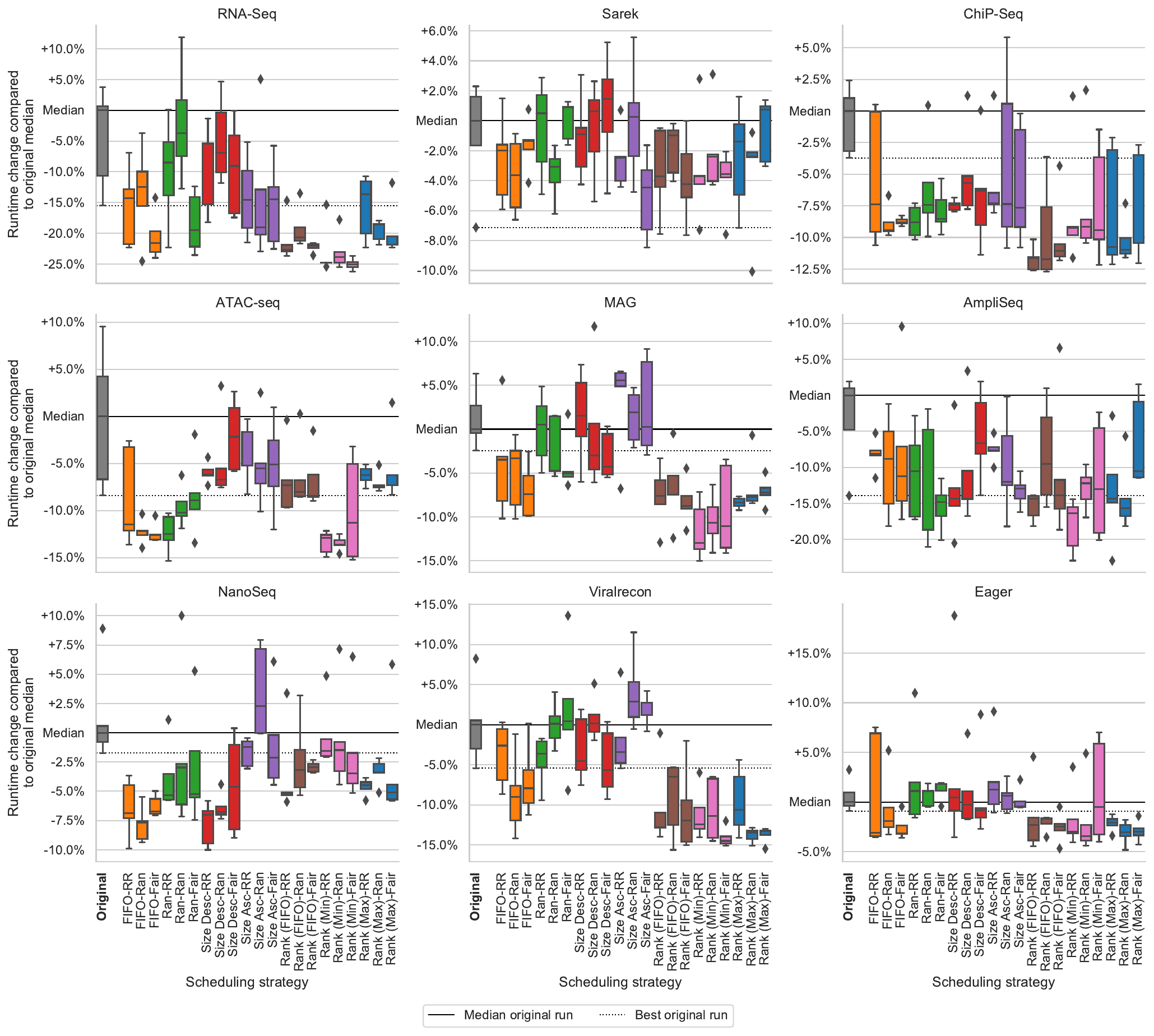}
  \caption{Runtime change by strategy and workflow compared to the median run of Nextflow's original strategy (left, grey)}
  \label{fig:runtime_change}
\end{figure*}

In our evaluation, we compare the original Nextflow-Kubernetes setup using the default Kubernetes scheduler with 21 scheduling strategies that we implemented for our scheduler.
\subsection{Experimental Setup}
We implemented seven prioritization strategies.
These strategies determine which task we assign to a node first.
We provide the following ordering strategies: Random, FIFO, descending and ascending input size, and task rank.
The rank is the number of following abstract tasks on the longest path.
Since several physical tasks can have the same rank, we implemented the rank strategy along with three different breakup strategies.
First, we apply FIFO for tasks with the same rank.
Second, we prefer tasks with larger input sizes, and third, we prefer tasks with smaller input sizes.

For node selection, we have three strategies.
We implemented Random, Round-robin, and Fair as node assignment strategies. 
Next, we combine each prioritization strategy with each node assignment strategy resulting in 21 different scheduling approaches.

We executed all 21 and the original strategy five times using nine of the eleven highest-ranked (number of GitHub stars) nf-core~\citep{ewelsNfcoreFrameworkCommunitycurated2020} workflows.
We take five runs because the Kubernetes startup overhead adds a high variance and the task runtime is relatively short.
Even with more runs, we would not be able to minimize the variance.
We publish the execution scripts and all collected raw data on GitHub\cref{git::experiments}.
We briefly introduce how to run the experiments in the Appendix. 
nf-core provides test data sets for each workflow, which we use accordingly.
Further, we adapted the workflow so that the data were available on the cluster out of the box to reduce the overhead of downloading data during the experiments.
For all workflows, we targeted to run for at least ten minutes for reliable measurements.
We skipped the MethylSeq and the RNA Fusion workflows because they failed with the test data sets provided.
Namely, we use the RNA-Seq, Sarek, ChiP-Seq, ATAC-seq, MAG, AmpliSeq, NanoSeq, Viralrecon, and Eager workflows.

We ran our experiments on a five-node cluster with 32 CPU cores and 128 GB of memory each, connected via a ten-gigabit network.
In the cluster, our scheduler and the Nextflow management pod run on the controller node, which was not used for task scheduling to avoid interference.

\begingroup 
\setlength{\tabcolsep}{0.48mm}
\begin{table*} \centering
    \rowcolors{2}{lightgray}{white}
    \captionsetup{singlelinecheck=off}
\caption[Aggregated values over all runs' runtime by strategy compared to original in percent]{\centering Aggregated values over all runs' runtime by strategy compared to original in percent; best value in bold.
    \small
  \begin{itemize}[leftmargin=2cm]
    \item \textit{Better med.}: How often are runs better than the original median
    \item \textit{Better min.}: How often are runs better than the best original run
    \item \textit{Med. better med.}: How often is the median run better than the original median run
    \item \textit{Med. change (avg.)}: Average runtime difference compared to the original median run
    \item \textit{Min change (avg.)}: Average runtime difference compared to the best original run
    \item \textit{Med. med. change (avg.)}: Average runtime difference compared between the strategy's and the original's median
    \item \textit{Med. med. change (best)}: Biggest improvement of the strategy's and the original's median
    \item \textit{Med. med. change (worst)}: Smallest improvement of the strategy's and the original's median
    \item \textit{Worst diff to worst ori.}: Difference between the strategy's and the original's worst run
    \item \textit{Max impr. to best ori.}: Difference between the strategy's and the original's best run
    \item \textit{Standard dev. (avg.)}: Average standard deviation over all workflows
    \item \textit{Standard dev. (best)}: Smallest standard deviation over all workflows
    \item \textit{Standard dev. (worst)}: Highest standard deviation over all workflows
  \end{itemize}}
    \begin{tabular}{c*{21}{|r}}
\multicolumn{1}{c}{} & \multicolumn{21}{c}{\textbf{Strategies}} \\
\multicolumn{1}{c}{Metric} & \multicolumn{1}{c}{\rot{FIFO-RR}} & \multicolumn{1}{c}{\rot{FIFO-Ran}} & \multicolumn{1}{c}{\rot{FIFO-Fair}} & \multicolumn{1}{c}{\rot{Ran-RR}} & \multicolumn{1}{c}{\rot{Ran-Ran}} & \multicolumn{1}{c}{\rot{Ran-Fair}} & \multicolumn{1}{c}{\rot{Size Desc-RR}} & \multicolumn{1}{c}{\rot{Size Desc-Ran}} & \multicolumn{1}{c}{\rot{Size Desc-Fair}} & \multicolumn{1}{c}{\rot{Size Asc-RR}} & \multicolumn{1}{c}{\rot{Size Asc-Ran}} & \multicolumn{1}{c}{\rot{Size Asc-Fair}} & \multicolumn{1}{c}{\rot{Rank (FIFO)-RR}} & \multicolumn{1}{c}{\rot{Rank (FIFO)-Ran}} & \multicolumn{1}{c}{\rot{Rank (FIFO)-Fair}} & \multicolumn{1}{c}{\rot{Rank (Min)-RR}} & \multicolumn{1}{c}{\rot{Rank (Min)-Ran}} & \multicolumn{1}{c}{\rot{Rank (Min)-Fair}} & \multicolumn{1}{c}{\rot{Rank (Max)-RR}} & \multicolumn{1}{c}{\rot{Rank (Max)-Ran}} & \multicolumn{1}{c}{\rot{Rank (Max)-Fair}} \\
\hline
Better med. & 86.7 & 97.8 & 93.3 & 75.6 & 73.3 & 73.3 & 80.0 & 68.9 & 75.6 & 77.8 & 53.3 & 75.6 & 95.6 & 93.3 & 97.8 & 91.1 & 91.1 & 93.3 & 97.8 & \textbf{100.0} & 86.7 \\
Better min. & 48.9 & 66.7 & 75.6 & 48.9 & 40.0 & 48.9 & 40.0 & 37.8 & 37.8 & 24.4 & 20.0 & 31.1 & 71.1 & 57.8 & 75.6 & \textbf{77.8} & 71.1 & 64.4 & 62.2 & 75.6 & 55.6 \\
Med. better med. & \textbf{100.0} & \textbf{100.0} & \textbf{100.0} & 66.7 & 77.8 & 66.7 & 77.8 & 77.8 & 88.9 & 77.8 & 44.4 & 77.8 & \textbf{100.0} & \textbf{100.0} & \textbf{100.0} & \textbf{100.0} & \textbf{100.0} & \textbf{100.0} & \textbf{100.0} & \textbf{100.0} & 88.9 \\
Med. change (avg.) & -6.0 & -7.8 & -8.2 & -5.4 & -4.2 & -6.0 & -4.7 & -3.4 & -3.8 & -3.5 & -2.6 & -4.6 & -9.1 & -7.3 & -8.6 & \textbf{-10.0} & -9.1 & -9.1 & -7.8 & -9.0 & -7.3 \\
Min change (avg.) & 0.8 & -1.1 & -1.7 & 1.4 & 2.7 & 0.6 & 2.1 & 3.6 & 3.1 & 3.3 & 4.2 & 2.1 & -2.7 & -0.7 & -2.2 & \textbf{-3.8} & -2.8 & -2.8 & -1.2 & -2.6 & -0.6 \\
Med. med. change (avg.) & -6.6 & -7.6 & -9.0 & -5.2 & -5.5 & -6.6 & -4.9 & -4.8 & -4.2 & -3.9 & -4.0 & -4.9 & -9.7 & -7.7 & -9.5 & \textbf{-10.8} & -9.8 & -10.2 & -8.0 & -9.2 & -7.7 \\
Med. med. change (best) & -14.3 & -12.5 & -21.6 & -12.5 & -18.7 & -19.5 & -14.4 & -14.3 & -9.1 & -14.6 & -19.0 & -14.5 & -22.7 & -20.6 & -22.0 & -24.8 & -23.9 & \textbf{-25.1} & -14.4 & -18.6 & -20.5 \\
Med. med. change (worst) & -2.0 & -1.9 & -1.3 & 1.1 & 1.1 & 1.2 & 1.5 & 0.6 & 1.5 & 5.6 & 2.9 & 2.7 & -2.3 & -1.0 & \textbf{-2.4} & -1.6 & -1.5 & -0.5 & -1.4 & -2.4 & 0.7 \\
Worst diff to worst ori. & 4.1 & 1.9 & 7.5 & 7.5 & 7.8 & 5.0 & 15.0 & 5.1 & 5.4 & 5.7 & 3.3 & 2.6 & 1.3 & -0.9 & 4.6 & 0.5 & 1.6 & 3.6 & -0.7 & \textbf{-3.0} & -0.4 \\
Max impr. to best ori. & -8.3 & -10.7 & -10.1 & -8.1 & -8.3 & -9.5 & -8.4 & -5.7 & -8.0 & -7.1 & -8.8 & -8.4 & -10.7 & -10.8 & -10.2 & \textbf{-12.9} & -11.9 & -12.7 & -10.5 & -10.2 & -10.6 \\
Standard dev. (avg.) & 4.5 & 3.8 & 3.0 & 4.1 & 4.5 & 3.6 & 4.3 & 4.6 & 4.6 & 3.7 & 5.2 & 3.8 & 3.3 & 3.7 & 3.3 & 3.4 & 3.3 & 4.2 & 3.1 & \textbf{2.2} & 3.3 \\
Standard dev. (best) & 2.2 & 1.3 & \textbf{0.3} & 1.3 & 1.0 & 0.9 & 0.4 & 1.2 & 2.7 & 1.2 & 1.5 & 1.2 & 1.0 & 0.8 & 0.5 & 1.2 & 0.8 & 0.9 & 0.6 & 0.9 & 1.0 \\
Standard dev. (worst) & 6.5 & 7.7 & 10.6 & 8.6 & 9.4 & 7.9 & 8.9 & 8.1 & 7.7 & 6.7 & 11.3 & 6.9 & 5.3 & 7.0 & 10.0 & 5.1 & 5.1 & 8.2 & 7.3 & \textbf{4.9} & 6.3 \\
\hline
\end{tabular}
    \label{tab::strategy_change}
\end{table*}
\endgroup
\subsection{Experimental Results}
Table~\ref{tab::strategy_runtimes} shows the characteristics of the workflow and gives an overview of the experiment.
In particular, the table shows the number of task instances, the data generated, the best strategy, the runtimes for the original and the best run, the median and the average task runtimes, the standard deviation over all task runtimes, and the improvement between the original and the best strategy for each workflow.
The generated data varies from 267.5 MB for AmpliSeq to 18.1 GB for the MAG workflow.

Table~\ref{tab::strategy_change} shows aggregated values over the runtimes of all runs and strategies.
The best measured values are indicated in bold. 
The best-performing prioritization strategies are the three rank strategies.
The average median-median difference over all nine workflows is at least 7.7\% better for all three rank strategies.
In particular, the Rank (Min) Round-robin strategy is better than the best original run in 77.8\% of the runs.
The Rank (Max) Random strategy is better than the median original run for all runs.
11 of the 21 strategies had for all nine workflows a better median than the original median.
Next, the table shows the average difference to the original's median (Med. change (avg.)).
Again, Rank (Min) Round-robin performs best, with an average runtime reduction of 10.0\% compared to the original's median and an average reduction of 3.8\% in runtime compared to the original's best value.

Figure~\ref{fig:runtime_change} shows the distribution of the percentage change in runtime for all nine workflows in comparison to the median runtime of Nextflow's original strategy.
The runtime is the time delta between the first and the last timestamp in Nextflow's log file.
The run using Nextflow's original strategy is highlighted in gray on the left.
The horizontal lines represent the median and minimum runtime measured for the original strategy within the five workflow repetitions.
Finally, we highlight all runs with the same prioritization strategy in the same color.

The plots show a vast variance across all workflows and strategies.
This is due to our small sample size and the very short runtime of the tasks.
Hence, Kubernetes' initialization time for each task plays a significant role.
In particular, the Random assignment strategies have the highest average variance.
For prioritization, the average variance of the max size strategy is slightly higher than the Random prioritization.
Also, the min size strategy has a significantly higher average variance than all other strategies.
Since the sample size does not vary much, ordering by size alone is more or less random.
The lowest variance is achieved by the Round-robin and Rank (Min) strategy.

For the Sarek workflow, all strategies perform equally.
This is because the workflow has the lowest number of tasks and one extremely long-running task that accounts for 80.8\% of the total runtime on average.

The FIFO Round-robin strategy is the closest to the original strategy.
However, the median runtime is always better than the median runtime of the original strategy.
We see two possible reasons for this. 
First, Kubernetes only does Round-robin fashion-like scheduling, and second, Kubernetes might consider many more node affinities.

In particular, Fair assignments for Rank (Min) and Rank (Max) often yield a much higher variance than their counterparts.
This is again due to the very small samples and short runtimes, often less than a second.
Kubernetes prepares each pod sequentially.
Accordingly, initialization becomes a significant part of the runtime for short-running tasks.
Round-robin and Random distribute the initialization time better over the nodes than Fair because Fair compensates one resource-demanding task on one node with multiple small tasks on another node.

Finally, we measure the overhead of our API.
Therefore, we calculate the difference between the median total runtime and the median makespan (the difference between the first task submitted and the last task completed).
On average, the median initialization difference for Nextflow was 2.7s, with a maximum of 5.1s using our API.
In contrast, the median total runtime was reduced by 71.3s on average, 160.0s in the best case, and at least 9.9s in the worst case.
Since this value is fixed, but the workflow's runtime is reduced by up to 25.1\%, we argue that the overhead is negligible. 

\section{Discussion}
In this section, we will first discuss the current limitations of our prototype.
Second, we will look at SWMSs other than Nextflow and resource managers other than Kubernetes to outline how to implement our API.
Finally, we give a brief outlook on further improvements to the scheduler in the future.

\subsection{Limitations}

One limitation is the additional consumed resources because the scheduler runs as an individual pod, requesting CPU and memory.
In our experiments, we did not lose any computing resources because the scheduler ran on a node not used for task execution to avoid interference.
Accordingly, we had the same memory and CPU available for all strategies.
However, management logic often runs on compute nodes.
But the resources used for the scheduler become negligible as the cluster size increases.

The scheduling strategies used in our prototype are very rudimentary. 
Prioritization and the node assignment work independently. 
As we used a homogeneous cluster, this is not an issue. 
However, for heterogeneous clusters, prioritization and node assignment should go hand in hand.
The necessary information is already available through our API.

Finally, we see barriers to bringing our API into existing resource managers and scientific management systems. 
It is an additional effort and, to some extent, a chicken-and-egg problem, as support on one side only makes sense if the other side also supports the API.
As we have shown the benefits of workflow-aware scheduling, implementing the API could increase interest and reduce the effort to support a particular system.
Further, we also see the possibility that some resource managers implement different algorithms than other resource managers, thereby gaining the interest of the workflow community.

\subsection{Usage with Other SWMSs}
In the following, we will discuss the effort and ability to extend SWMSs.
We will discuss this for three well-known SWMSs.
\paragraph{Snakemake} uses the official Python 3 Kubernetes library for interaction. 
Therefore, one can configure the target scheduler, resource limits, and other additional information. 
Similar to Nextflow, Snakemake loops over the ready-to-run tasks. 
Again, a batching mechanism could be integrated at this point, and Snakemake could submit the task information.
Further, Snakemake provides a DAG that can be transferred initially.
\paragraph{Airflow} also uses the official Python 3 Kubernetes library. 
Airflow uses a heartbeat that triggers the creation of new jobs and internally uses a queue system for ready-to-run tasks. 
Here, the batching mechanism could be integrated, and the task information could be transferred.
One difference to the previously presented SWMSs is the mandatory SQL database where Airflow reads the DAG from and could send it to our API.
\paragraph{Argo} uses the Kubernetes client written in Go. 
The internal operator executes the DAG and identifies the respective ready-to-run tasks. 
Here, Argo could pass additional metadata to our proposed API. 
This step is followed by iterating over the ready-to-run tasks where batching could be applied.

\subsection{Usage with Other Resource Managers}
Extending the resource managers requires a deeper dive into the code base.
Especially making our scheduler the only responsible component for scheduling a task is of interest.

\paragraph{Slurm~\cite{slurm}} can be extended with plugins.
For example, Slurm already offers a REST-API plugin\footnote{\url{https://slurm.schedmd.com/rest.html}}.
In particular, this plugin can be extended with our proposed REST-API endpoints to transfer information.
Further, Slurm has a central database to store task information.
Here, we can add workflow-specific data.

Compared to our Kubernetes scheduler, scheduling in Slurm is divided into two independent tasks.
First, Slurm prioritizes the tasks using a scheduler plugin\footnote{\url{https://elwe.rhrk.uni-kl.de/documentation/schedplugins.html}}.
We can extend this to consider the DAG structure and apply workflow scheduling algorithms.
Second, the node selector plugin\footnote{\url{https://slurm.schedmd.com/select_design.html}} is responsible for assigning a task to a concrete machine.
Therefore, scheduling strategies, as shown in this paper, are possible.

\paragraph{HTCondor~\citep{SterlingCondor}} uses bidding logic to assign nodes to tasks.
Accordingly, we must fundamentally change this logic to implement our own strategies.
HTCondor allows the user to adapt task priorities.
Further, HTCondor offers DAGMan, an extension to schedule DAG structured workflows.
To the best of our knowledge, HTCondor does not offer plugins for extensibility.
DAGMan only supports static workflows, so we would need another external tool to provide the API, wrap the tasks for HTCondor, and take care of the priorities.
However, this approach would not support node assignments.
Thus, modifying node assignments would require significant changes in HTCondor's code base.

\paragraph{Hadoop} uses YARN~\citep{vavilapalliApacheHadoopYARN2013a} to schedule and assign tasks to nodes.
YARN provides ApplicationMasters to coordinate applications.
These ApplicationMasters run as a regular container, comparable to our Kubernetes scheduler.
Hi-Way~\citep{buxHiWAYExecutionScientific2017}, for example, uses this concept to implement workflow-aware scheduling algorithms in Hadoop.
To bring our approach to Hadoop, we could implement our own ApplicationMaster or extend Hi-Way with our REST-API and use the data accordingly.
Since YARN plugins can be written in Java, it is possible to reuse most of our Kubernetes implementation.

Today, Nextflow does not support Hadoop because Hadoop uses HDFS for file storage, which is not POSIX compliant.
However, with task input information, Nextflow could work with Hadoop as the ApplicationMaster could stage input files from the HDFS before and after workflow execution.
Hi-Way already does this to enable black-box task execution.

\section{Conclusion}
In this paper, we developed a common scheduling interface between SWMSs and resource managers.
First, we extracted the information needed for workflow-aware scheduling.
In the next step, we designed an API to exchange this information between an SWMS and a resource manager.
With this API, we effectively decouple the SWMS from a concrete resource manager.
Therefore, we designed a common API that can simplify access to various resource managers.

In our experiments, we tested our approach for making the resource manager's scheduler workflow-aware by using Nextflow and Kubernetes.
We showed that using our API interface to exchange information between SWMSs and resource managers significantly reduces the makespan.
We measured nine different nf-core workflows with 21 scheduling strategies in comparison to the default strategy. 
In our experiments, we improved the median runtime by up to 25.1\%, while the overhead of the information transfer remains negligible.
The best-performing strategy showed an average makespan reduction of 10.8\%.
In our experiments, the more informed rank strategies significantly outperform FIFO, random, and ordering by the input size.

In the future, we plan to include more sophisticated scheduling strategies that can take full advantage of the information provided by our common API.
The scheduling algorithms do not necessarily have to optimize for makespan, other objectives are also possible.
Integrating a provenance storage and a provenance API would further leverage the potential of the scheduler interface for more advanced scheduling.
The prototype can be a first step toward making scheduling configurable by the user and for scheduling scientists to validate their scheduling algorithm with real-world workflows and systems~\citep{osti_10358034}.

\section*{Acknowledgment}
We thank Rafael Ferreira da Silva for his valuable feedback and input on our initial version of this paper.

This work was funded by the German Research Foundation (DFG), CRC 1404: "FONDA: Foundations of Workflows for Large-Scale Scientific Data Analysis."

\bibliographystyle{IEEEtranN}
\bibliography{bibliography}

\clearpage
\appendix

\section{Artifact Description Appendix: [Paper Title]}

\subsection{Abstract}

In this artifact, we describe how to build the Common Workflow Scheduler (CWS) for Kubernetes and the adapted Nextflow version proposed in the paper: ``How Workflow Engines Should Talk to Resource Managers: A Proposal for a Common Workflow Scheduling Interface''.
Moreover, we describe the experimental setup and how to prepare and get the input data of the nine presented workflows and run them with the 22 scheduling strategies.

\subsection{Description}

\subsubsection{Check-list (artifact meta information)}

{\small
\begin{itemize}
  \item {\bf Algorithm: } All seven prioritizing and three assignment strategies are implemented in the Common Workflow Scheduler (CWS) for Kubernetes.
  \item {\bf Program: } CWS for Kubernetes~\cite{lehmannCWSk8s} and Nextflow with CWS extension for Kubernetes~\cite{lehmannCWSnextflow}.
  \item {\bf Compilation: } Nextflow with CWS extension for Kubernetes and Common Workflow Scheduler for Kubernetes as Docker image.
  \item {\bf Data set: } Scripts to download the input data. Traces and logs of all 990 workflow executions~\cite{lehmannCWSeval}.
  \item {\bf Run-time environment: } Kubernetes cluster in version \emph{1.23.6}.
  \item {\bf Hardware: } All nodes were x86-64 machines.
  \item {\bf Runtime state: } The cluster was exclusively used for the experiments.
  \item {\bf Execution: } Bash scripts to manage the experiment, Dockerfiles and scripts to build the two programs.
  \item {\bf Output: } Workflow traces and logs of all 990 workflow executions that are processed to generate the tables and plots in the paper.
  \item {\bf Experiment workflow: } Download the inputs, and run the execution file.
  \item {\bf Experiment customization: } It is possible to use different/additional/repeated data samples or to add another workflow to the experiment.
  \item {\bf Publicly available?: } Yes, all code and experimental results are hosted by us. Input data is obtained from public sources.
\end{itemize}
}

\subsubsection{How software can be obtained}
The Common Workflow Scheduler for Kubernetes implementation can be cloned from GitHub~\href{https://github.com/CommonWorkflowScheduler/KubernetesScheduler}{https://github.com/CommonWorkflowScheduler /KubernetesScheduler}.
Moreover, we provide an already-built Docker image (commonworkflowscheduler/kubernetesscheduler:v1.0) starting the Common Workflow Scheduler service.

The adapted Nextflow version can be cloned from GitHub~\href{https://github.com/CommonWorkflowScheduler/Nextflow}{https://github.com/CommonWorkflowScheduler /Nextflow}.
Again, a prebuilt Docker image is available on DockerHub (commonworkflowscheduler/nextflow:v1.0).

\subsubsection{Hardware dependencies}
All artifacts are tested and prebuilt with x86-64 machines running Ubuntu \emph{20.04}. 

\subsubsection{Software dependencies}
Building the software requires Docker and Java OpenJDK to be installed.
CWS uses Maven as a build system. 
Maven only needs to be installed if CWS is not built using Docker.
Nextflow uses Gradle as a build system and make - Gradle does not need to be installed.
Moreover, a running Kubernetes Cluster is required.
Alternatively, a local Kind or Minikube setup will work as well.
For our experiments, we used Ceph as a shared filesystem. 
However, any read-write-many filesystem supported by Nextflow will work.
We need kubectl installed to communicate with the cluster and orchestrate the experiment.
Moreover, Nextflow should be in the PATH.

\subsubsection{Data sets}
For all nine workflows, we use the data set provided in the test profile that all nf-core workflows offer.
To extend the runtime, we have partially extended these data sets with data from other profiles or included the same data multiple times.
Input data is prefetched to avoid affecting experiment runtimes with download times.
The download is performed by executing the \emph{fetchdata.sh} file provided for all workflows in~\href{https://github.com/CommonWorkflowScheduler/ExperimentsAndResults/tree/main/inputs}{https://github.com/CommonWorkflowScheduler /ExperimentsAndResults/tree/main/inputs}.
Moreover, we provide a configuration file for each workflow.

\subsection{Installation}

\subsubsection{Build Nextflow}
Before we build Nextflow, we have to adjust two files.
First, in Nextflow's root directory in the \emph{nextflow} file, we add a ``return'' into the first line of the ``get'' method.
Otherwise, our build will automatically update, downloading the newest official Nextflow version.
Second, in the \emph{docker} folder, the \emph{Makefile} needs adjustments.
Change the first line of the \emph{build} target from \small\mintinline{bash}{cp ../nextflow .} to \small{\mintinline{bash}{cp ../build/releases/nextflow*-all nextflow}}.
This is necessary, as we cannot download the jar file.
Then replace the body of the \emph{release} target with the following two lines. 
Use your docker name and define an arbitrary version tag.
\begin{lstlisting}[float=ht,language=bash]
docker tag nextflow/nextflow:${version} <your docker account>/cws:<version>
docker push <your docker account>/cws:<version>
\end{lstlisting}

Run the following instructions to build and publish the Nextflow with CWS Docker image.
\begin{lstlisting}[float=ht,language=bash]
$ cd <nextflow root directory>
# change the nextflow file as described
$ make compile
$ make pack
$ make install
$ cd ./docker
# change the Makefile as described
# login to Docker
$ make release
\end{lstlisting}

\subsubsection{Build Common Workflow Scheduler}
To build the CWS, run the following commands.
\begin{lstlisting}[float=ht,language=bash]
$ cd <CWS root directory>
$ docker build -t cws .
$ docker tag cws <your docker account>/cws:<version>
$ docker push <your docker account>/cws:<version>
\end{lstlisting}

\subsubsection{Prepare the cluster}
To prepare the cluster, download the following project: \hspace{4mm}\href{https://github.com/CommonWorkflowScheduler/ExperimentsAndResults}{https://github.com/CommonWorkflowScheduler /ExperimentsAndResults}.
In the \emph{setup} directory, adjust the \emph{pvc.yaml} file.
Set the storageClassName to a storage class installed in your cluster.
If you change the experimental setup, you might also change the storage request.
By default, the experiments run in the \emph{cws} namespace.
To define another namespace, adjust the namespace in the \emph{setup.sh} file and in \emph{setup/accounts.yaml} (two times).
Afterward, run the \emph{setup.sh} file in the ExperimentsAndResults root directory \small{\mintinline{bash}{bash setup.sh}}.
This script will download the data sets and workflows and prepare the cluster for the experiments.

\subsubsection{Local experiment management environment}
We highly recommend running the experiment from a node within the cluster to avoid connection breaks.
First, we limit the number of nodes used in order to have a scheduling problem. 
Therefore, we label the nodes executing the tasks and the node running the scheduler. 
Both should be distinct.
The number of nodes should be much smaller than the parallel tasks executed.
In our experiments, we used four nodes with 32 cores each.
To label the nodes, run the following:
\begin{lstlisting}[float=ht,language=bash]
# Nodes executing tasks
kubectl label nodes <Node 1> <Node 2> cwsexperiment=true
# Node that runs the scheduler
kubectl label nodes <Node 0> cwsscheduler=true
\end{lstlisting}

Next, in the \emph{nextflow.config} file in the \emph{execution} directory, set the namespace to the namespace used before (two times).
Moreover, change the namespace in the \emph{runExperiments.sh}.
The experiment will create the \emph{evaluation} folder. 
Accordingly, you can delete the folder in the Git project to store new measurements.
Existing measurements are skipped.

\subsection{Experiment Workflow}
The experiment is completely orchestrated by a bash script.
To launch the experiment:
\begin{lstlisting}[float=ht,language=bash]
cd <root directory of ExperimentsResults>
cd execution
# The name could be the cluster to run.
bash runExperiments.sh <name>
\end{lstlisting}

The script will do the following:
\begin{enumerate}
    \item load all required images into Kubernetes' cache on all nodes
    \item run the workflow for all strategies
    \item download the results
\end{enumerate}

\subsection{Evaluation and Expected Result}

All our measurements can be found in the \emph{evaluation/CPU} folder \href{https://github.com/CommonWorkflowScheduler/ExperimentsAndResults}{https://github.com/CommonWorkflowScheduler /ExperimentsAndResults}.
In our group, we call the experiment cluster the CPU Cluster. 
For each of the nine workflows, 22 Strategies with five runs each are executed.
For each run, we collect the trace file, the logs from the Nextflow and the CWS console, the Nextflow logs, the used config, the generated report, and the workflow's dag.
Besides the measurements, we provide the Jupyter script (\emph{Evaluation.ipynb}) that generated the plots and tables in the paper.

Different runtimes for the different strategies are expected.
If this is not the case, the setup might use too many nodes.
All tasks can start simultaneously in such a setup, and we do not have a scheduling problem.
Accordingly, the choice of algorithm does not make a difference, as no algorithm can prioritize tasks better.

\subsection{Experiment Customization}
The data sets can be changed or other workflows included.
Therefore, Nextflow's config has to be adjusted.
Moreover, the storage capacity should be increased.

\end{document}